\documentclass[showpacs,amsmath,amssymb,aps,prd,superscriptaddress,nofootinbib,twocolumn]{revtex4}

\usepackage{epstopdf}
\usepackage{graphicx}
\usepackage{amsmath}
\usepackage{multirow}
\usepackage{latexsym}

\def\Gammaflat{\hat \Gamma}

\def\Dflat{\hat {\mathcal D}}
\def\part_n{\partial_\perp}

\def\4{{}^{(4)}}

\begin{document}
   
\title{General relativistic hydrodynamics in curvilinear coordinates}

\author{Pedro J. Montero}
\affiliation{Max-Planck-Institut f{\"u}r Astrophysik,
Karl-Schwarzschild-Str.~1, D-85748, Garching bei M{\"u}nchen, Germany}

\author{Thomas W. Baumgarte}
\affiliation{Max-Planck-Institut f{\"u}r Astrophysik,
Karl-Schwarzschild-Str.~1, D-85748, Garching bei M{\"u}nchen, Germany}
\affiliation{Department of Physics and Astronomy, Bowdoin College, Brunswick, ME 04011, USA}

\author{Ewald M\"{u}ller}
\affiliation{Max-Planck-Institut f{\"u}r Astrophysik,
Karl-Schwarzschild-Str.~1, D-85748, Garching bei M{\"u}nchen, Germany}

\begin{abstract}
 In this paper we report on what we believe is the first successful implementation of relativistic hydrodynamics, coupled to dynamical spacetimes, in spherical polar coordinates without symmetry assumptions.  We employ a high-resolution shock-capturing scheme, which requires that the equations be cast in flux-conservative form.  One example of such a form is the ``Valencia" formulation, which has been adopted in numerous applications, in particular in Cartesian coordinates.  Here we generalize this formulation to allow for a reference-metric approach, which provides a natural framework for calculations in curvilinear coordinates.  In spherical polar coordinates, for example, it allows for an analytical treatment of the singular $r$ and $\sin\theta$ terms that appear in the equations.  We experiment with different versions of our generalized Valencia formulation in numerical implementations of relativistic hydrodynamics for both fixed and dynamical spacetimes.  We consider a number of different tests -- non-rotating and rotating relativistic stars, as well as gravitational collapse to a black hole -- to demonstrate that our formulation provides a promising approach to performing fully relativistic astrophysics simulations in spherical polar coordinates.

\end{abstract}

% \pacs{04.25.dg, 04.70.Bw, 97.60.Jd, 97.60.Lf}

\maketitle

%==========================================================
\section{Introduction}
%==========================================================

Solving many problems of great astrophysical interest, including gravitational collapse to black holes, mergers of a neutron stars with a binary companion, black-hole accretion disks, and supernovae explosions, requires modeling relativistic fluids in either fixed or dynamical spacetimes.   A key feature of inviscid fluids is the appearance of shocks and contact discontinuities, i.e.~the development of discontinuities in the fluid variables.   These discontinuities pose a challenge to traditional numerical methods, for example finite-difference or spectral methods, so that special numerical algorithms have been developed for fluid dynamics.

Many recent applications employ so-called high-resolution shock-capturing (HRSC) methods \cite{God59,HarLL83,Ein88}.  At the core of these methods are Riemann solvers that produce either exact or approximate solutions to Riemann problems and use these solutions to update the fluid variables in each grid cell (see, e.g., \cite{Tor99,Lev92} for an introduction; see also \cite{MarM99,Fon00} for reviews.)   The application of such HRSC methods requires that the equations of hydrodynamics  be cast in flux-conservative form.   A commonly used flux-conservative form of the equations of relativistic hydrodynamics is the so-called ``Valencia" formulation \cite{BanFIMM97}.  This form of the equations has been used successfully in a large number of simulations.  Some of these simulations hold the spacetime fixed, others adopt some approximation method to evolve the gravitational fields, while others yet evolve the relativistic gravitational fields self-consistently together with the fluid.

To date, most self-consistent calculations that do not make any symmetry assumptions adopt Cartesian coordinates (see~\cite{FonMST00,FonGIMRSSST02,ShiF05,BaiHMLRSFS05,NeiHM06,MonFS08,EtiFLSTB08,AndHirLeh08,ThiBB10,Lofetal12} for some examples; see also \cite{Reis13,Ottetal13} for a multi-patch implementation, and the CoCoNut code \cite{coconut} for an implementation in spherical polar coordinates using a conformal-flatness approximation for the gravitational fields.)  While Cartesian coordinates have some desirable properties for some applications,  other applications benefit from spherical polar or other curvilinear coordinates. Specific examples include simulations of gravitational collapse, supernovae, and accretion disks. 

In \cite{BauMCM13} we have recently introduced a new approach for the
evolution of gravitational fields in spherical polar coordinates.  Our
method adopts the Baumgarte-Shapiro-Shibata-Nakamura (BSSN)
formulation \cite{NakOK87,ShiN95,BauS98} in a covariant,
reference-metric approach  \cite{Gou07,Bro09,Gou12} (see also \cite{BonGGN04,ShiUF04,CooB08}) and evolves the resulting equations using a partially implicit Runge-Kutta (PIRK) time integration method \cite{MonC12,CorCD12}.  The reference-metric approach introduces several desirable features; in particular it plays a crucial role in casting the equations in a form that allows for an analytical treatment of the singular $r$ and $\sin\theta$ terms that appear in spherical polar coordinates.  This analytical treatment of the singular terms, in turn, allows for stable numerical simulations without the need to regularize the equations.

 The purpose of this paper is twofold.  We first generalize the
 Valencia formulation of relativistic hydrodynamics to allow for such
 a reference-metric.  An attractive feature of this generalization,
 besides the fact that the resulting equations mesh well with those
 for the gravitational fields expressed in a reference-metric
 approach, is that all hydrodynamical quantities, their fluxes and
 source terms, are now defined as proper tensorial quantities (of
 weight zero).  We derive this formalism in general and without
 specializing to any coordinate system, but highlight some specific
 advantages of the reference-metric approach for numerical simulations
 in spherical polar coordinates.  We then report on successful
 numerical implementations of these equations coupled to Einstein's
 equations for the gravitational fields, in three spatial dimensions,
 and without the need of regularization. We refer to~\cite{BauMCM13}
 for details of our approach for the evolution of Einstein's
 equations in spherical polar coordinates.\footnote{By ``regularization" we mean a reformulation of the equations in which all singular terms are eliminated with the help of a new set of dynamical variables.}  
 
 We experiment with different combinations of using the equations of hydrodynamics with and without the reference-metric approach and find that, while the reference-metric or some other accommodation of the spherical polar coordinates is indeed crucial in the Euler equation, numerical errors are smaller if the continuity and energy equation are left in the original version.  We perform several tests for non-rotating and rotating relativistic stars as well as collapse to black holes.   Our results demonstrate that our formulation and methods provide a promising approach to performing fully relativistic simulations in spherical polar coordinates, and that they are well-suited for future applications in simulations of supernovae, gravitational collapse and other objects of interest in relativistic astrophysics.

Our paper is organized as follows.  In Section \ref{Sec:3+1} we
briefly review the 3+1 decomposition of Einstein's field equations,
introduce the notion of a reference-metric, and present those
expressions that are needed in the rest of the paper.  In Section
\ref{Sec:ERH} we rederive the equations of relativistic hydrodynamics,
generalizing the approach of the Valencia formalism to allow for a
general reference metric.  We compare with the original Valencia
formalism and highlight advantages of our formalism in curvilinear
coordinates.  In Section \ref{Sec:SPC} we describe our numerical
implementation of these equations.  In Section \ref{Sec:examples} we
present numerical results in spherical polar coordinates; in
particular we show results for non-rotating and rotating relativistic
stars, with and without Cowling approximation \cite{Cow41},
Oppenheimer-Snyder dust collapse and the collapse of a marginally
stable static star to a black hole.  We briefly summarize our findings
in Section \ref{Sec:summary}.  Throughout this paper we adopt
geometric units in which $G = c = M_{\odot} = 1$. However, we express
time in milliseconds for the simulations of spherical and rotating stars, and for the Oppenheimer-Snyder dust collapse we
use units $G = c = 1$ to ease the comparison with the literature.

%==========================================================
\section{The 3+1 decomposition}
\label{Sec:3+1}
%==========================================================

We assume that the spacetime $M$ can be foliated by a family of spatial slices $\Sigma$ that coincide with level surfaces of a coordinate time $t$.   We denote the future-pointing unit normal on $\Sigma$ with $n^a$ and write the spacetime metric $g_{ab}$ as
\begin{eqnarray} \label{metric}
ds^2 & = & g_{ab} dx^a dx^b \nonumber \\ 
& = & - \alpha^2 dt^2 + \gamma_{ij} (dx^i + \beta^i dt)(dx^j + \beta^j dt),
\end{eqnarray}
where $\alpha$ is the lapse function, $\beta^i$ the shift vector, and $\gamma_{ij}$ the spatial metric induced on $\Sigma$,
\begin{equation} \label{spatial_metric_def}
\gamma_{ab} = g_{ab} + n_a n_b.
\end{equation}
Here and in the following indices $a, b, \ldots$ run over spacetime indices, while indices $i, j, \ldots$ run over space indices only.  In terms of the lapse and shift, the normal vector $n^a$ can be expressed as
\begin{equation} \label{normal}
n_a = (-\alpha,0,0,0)~~~\mbox{or}~~~n^a = (1/\alpha, - \beta^i/\alpha).
\end{equation}

We adopt a conformal decomposition of the spatial metric $\gamma_{ij}$ 
\begin{equation} \label{conformal_decomposition}
\gamma_{ij} = e^{4 \phi} \bar \gamma_{ij},
\end{equation}
where $\psi = e^\phi = (\gamma/\bar \gamma)^{1/12}$ is the conformal factor and $\bar \gamma_{ij}$ the conformally related metric.

For applications in curvilinear coordinates it is convenient to introduce a reference metric $\hat \gamma_{ij}$. We will specialize to spherical polar coordinates in Section \ref{Sec:SPC}, but for now the only assumption that we will make for $\hat \gamma_{ij}$ is that its determinant $\hat \gamma$ be independent of time.  Even this assumption would be easy to relax, for example for applications in cosmology.  

Associated with the different types of metrics are different covariant derivatives.  In the following we denote the covariant derivative associated with the spacetime metric $g_{ab}$ by $\nabla_a$, that associated with the spatial metric $\gamma_{ij}$ with $D_i$, the covariant derivative associated with the conformally related metric $\bar \gamma_{ij}$ with $\bar D_i$, and finally the covariant derivative associated with the reference metric $\hat \gamma_{ij}$ with $\Dflat_i$.  We also denote the corresponding connection symbols with $\4 \Gamma^a_{bc}$, $\Gamma^i_{jk}$, $\bar \Gamma^i_{jk}$ and $\Gammaflat^i_{jk}$, respectively.  We define
\begin{equation} \label{delta_Gamma}
\Delta \Gamma^i_{jk} \equiv \bar \Gamma^i_{jk} - \Gammaflat^i_{jk}
\end{equation}
and note that, unlike the connection symbols themselves, these differences are tensors, and that they can be computed from
\begin{equation} \label{delta_Gamma_compute}
\Delta \Gamma^i_{jk} = \frac{1}{2} \bar \gamma^{il} ( 
	\Dflat_j \bar \gamma_{lk} + \Dflat_k \bar \gamma_{lj} - \Dflat_l \bar \gamma_{jk} ).
\end{equation}
If the reference metric is chosen to be the flat metric in Cartesian coordinates, the covariant derivative $\Dflat_i$ reduces to the partial derivative $\partial_i$, all $\Gammaflat^i_{jk}$ vanish, and 
$\Delta \Gamma^i_{jk} = \bar \Gamma^i_{jk}$.

We assume that a numerical solution for the spacetime metric $g_{ab}$ is constructed by evolving the spatial metric $\gamma_{ij}$ forward in time. Such an evolution also involves the extrinsic curvature $K_{ij}$
\begin{equation} \label{K_def}
K_{ij} \equiv - \gamma_i{}^k \gamma_j{}^l \nabla_k n_l.
\end{equation}
The extrinsic curvature can also be expressed as 
\begin{equation}
K_{ij} = - \frac{1}{2 \alpha} \partial_t \gamma_{ij} + D_{(i} \beta_{j)},
\end{equation}
which highlights its role as the time derivative of the spatial metric.

%==========================================================
\section{Relativistic hydrodynamics with a reference metric}
\label{Sec:ERH}
%==========================================================

 The equations of relativistic hydrodynamics are based on conservation of rest mass, expressed by the continuity equation 
\begin{equation} \label{continuity}
\nabla_a (\rho_0 u^a) = 0,
\end{equation}
and conservation of energy-momentum,
\begin{equation} \label{div_T}
\nabla_b T^{ab} = 0.
\end{equation}
Here $\rho_0$ is the rest-mass density, $u^a$ the fluid four-velocity, and $T^{ab}$ the stress-energy tensor 
\begin{equation} \label{stress-energy}
T^{ab} = \rho_0 h u^a u^b + p g^{ab},
\end{equation}
where $h \equiv 1 + \epsilon + p/\rho_0$ is the enthalpy, $p$ the pressure, and where $\epsilon$ is the specific internal energy.  The quantities $\rho_0$, $p$, $\epsilon$ and the fluid velocity $v^i$ defined in equation (\ref{v}) below form the so-called {\em primitive} fluid variables.

In most recent applications, the above equations are brought into flux-conservative form, so that high-resolution shock-capturing (HRSC) schemes can be used to find accurate numerical solutions.  In the process, a new set of hydrodynamic variables, namely the {\em conserved} variables, are introduced.  An example of such a flux-conservative form is the ``Valencia" form of the equations (see, e.g., \cite{MarM99,ThiBB10}.)   While these equations are fully covariant, they are, in their original form, not yet well suited for applications in curvilinear coordinates, as we will explain in more detail below.  In the following we derive an alternative version of these equations that is based on a reference metric approach. In Section \ref{Sec:examples} we will experiment with numerical implementations of this new formulation, and will find that it has significant advantages in curvilinear coordinates, at least for the Euler equation derived in Section \ref{Sec:Euler}.

%==========================================================
\subsection{The continuity equation}
\label{Sec:Continuity}
%==========================================================

The covariant divergence of a vector $V^a$ can be expressed as
\begin{equation} \label{vec_ident}
\nabla_a V^a = \frac{1}{\sqrt{|g|}} \partial_a \left( \sqrt{|g|} \, V^a \right),
\end{equation}
(see, e.g., Problem 8.16 (c) in \cite{MisTW73}, or Problem 7.7 (g) in \cite{LigPPT75}), which holds for any metric and its associated covariant derivative.  In the following, we will use this identity twice; once for the spacetime metric $g_{ab}$, and once for the reference metric $\hat \gamma_{ij}$.  

We start by applying (\ref{vec_ident}) for the spacetime metric, for which $|g| = - g$, to the continuity equation (\ref{continuity}) to obtain
\begin{eqnarray} \label{cont1}
0 & = & \nabla_a (\rho_0 u^a) = \frac{1}{\sqrt{-g}} \partial_a \left( \sqrt{-g} \rho_0 u^a \right) \nonumber \\
& = & \frac{1}{\sqrt{-g}} \left( \partial_t \left( \sqrt{-g} \rho_0 u^t \right) + \partial_j \left( \sqrt{-g} \rho_0 u^j\right)  \right).
\end{eqnarray}
We now use eqs.~(\ref{metric}) and (\ref{conformal_decomposition}) to expand the determinant of the spacetime metric as
\begin{equation} \label{det}
\sqrt{-g} = \alpha \sqrt{\gamma} = \alpha e^{6 \phi} \sqrt{\bar \gamma}
\end{equation}
and write the spatial terms in (\ref{cont1}) as
\begin{eqnarray}
\partial_j \left( \alpha e^{6 \phi} \sqrt{\bar \gamma} \rho_0 u^j\right)  & = &  
\partial_j \left( \sqrt{\hat \gamma} \alpha e^{6 \phi} \sqrt{\bar \gamma/\hat \gamma} \, \rho_0 u^j\right)   \\
& = & \sqrt{\hat \gamma} \, \Dflat_j \left( \alpha e^{6 \phi} \sqrt{\bar \gamma/\hat \gamma} \, \rho_0 u^j\right).  \nonumber
\end{eqnarray}
Here we have used the identity (\ref{vec_ident}) for the reference metric $\hat \gamma_{ij}$ in the last step.  Inserting this last result into (\ref{cont1}) we obtain
\begin{equation} \label{D_cov}
\partial_t (e^{6 \phi} \sqrt{\bar \gamma/\hat \gamma} \, D) + \Dflat_j (f_D)^j = 0,
\end{equation}
where we have defined the density as seen by a normal observer
\begin{equation} \label{D}
D \equiv W \rho_0
\end{equation}
and the corresponding flux
\begin{equation} \label{f_D}
(f_D)^j \equiv \alpha e^{6 \phi} \sqrt{\bar \gamma/\hat \gamma} D (v^i - \beta^i/\alpha).
\end{equation}
Here
\begin{equation}
W \equiv - n_a u^a = \alpha u^t
\end{equation}
is the Lorentz factor between the fluid and a normal observer, and
\begin{equation} \label{v}
v^a \equiv \gamma^a{}_b \left( \frac{u^b}{W} + \frac{\beta^b}{\alpha} \right)
\end{equation}
is the fluid velocity as measured by a normal observer.  We note that we have assumed in eq.~(\ref{D_cov}) that $\hat \gamma$ is independent of time; as we said before, this could be generalized quite easily.

The form of (\ref{D_cov}) is exactly as in the original Valencia formulation, except for the appearance of the factors $\sqrt{\hat \gamma}$ in (\ref{D_cov}) and (\ref{f_D}), and the covariant derivative with respect to the reference metric, $\Dflat_j$, in (\ref{D_cov}).  Choosing a flat metric in Cartesian coordinates reduces the former to unity and the latter to a partial derivative, so that the corresponding equation in the Valencia formulation is recovered.   We also note that we can derive equation (\ref{D_cov}) from the corresponding Valencia equation directly by inserting a factor $1 = \sqrt{\hat \gamma}/\sqrt{\hat \gamma}$ into the flux term $(f_D)^i$, and then using the product rule for the partial derivative.

In a numerical implementation the covariant derivative in
(\ref{D_cov}) should be evaluated in terms of partial derivatives and
connection symbols (rather than the identity (\ref{vec_ident})).  Since $(f_D)^j$ is a tensor density of weight zero we obtain
\begin{equation} \label{D_partial}
\partial_t (e^{6 \phi} \sqrt{\bar \gamma/\hat \gamma} D) + \partial_j (f_D)^j = - (f_D)^j \Gammaflat^k_{jk}.
\end{equation}
Here the $\Gammaflat^k_{jk} = \partial_j \ln \sqrt{\hat \gamma}$ can be evaluated analytically from the known reference metric.

%==========================================================
\subsection{The Euler equation}
\label{Sec:Euler}
%==========================================================

The divergence of a mixed-index second-rank tensor $A_a{}^b$ can be expressed as
\begin{equation} \label{tens_ident}
\nabla_b A_a{}^b = \frac{1}{\sqrt{-g}} \, \partial_b \left( \sqrt{-g} \, A_a{}^b \right) - A_c{}^b \4\Gamma^c_{ba}
\end{equation}
(see Problem 7.7 (h) in \cite{LigPPT75}), which again holds for any metric and its associated covariant derivative.  

We now derive the Euler equation by applying (\ref{tens_ident}) for the spacetime metric to a spatial projection of equation (\ref{div_T}),
\begin{eqnarray} \label{div_T1}
0 & = &  \gamma_{ib} \nabla_a T^{ab} = g_{ib} \nabla_a T^{ab} = \nabla_a ( g_{ib}T^{ab} ) 
\nonumber \\
& = & \frac{1}{\sqrt{-g}} \partial_a \left( \sqrt{-g} T_i{}^a \right) - T_a{}^b \4 \Gamma^a_{ib}  \\
& = & \frac{1}{\sqrt{-g}} \left( \partial_t \left( \sqrt{-g} T_i{}^t \right)  + 
\partial_j \left( \sqrt{-g} T_i{}^j \right) \right) - T_a{}^b \4 \Gamma^a_{ib}, \nonumber 
\end{eqnarray}
Using (\ref{det}) we now expand
\begin{eqnarray}
&&\partial_j \left( \sqrt{-g} T_i{}^j \right) 
 = \partial_j \left( \sqrt{\hat \gamma } \, \alpha e^{6 \phi} \sqrt{\bar \gamma/\hat \gamma} T_i{}^j \right) \\
&& ~~~~~ =  \sqrt{\hat \gamma} \, \Dflat_j \left( \alpha e^{6 \phi} \sqrt{\bar \gamma/\hat \gamma} T_i{}^j \right)
+ \alpha e^{6 \phi} \sqrt{\bar \gamma} T_k{}^j \Gammaflat^k_{ij}, \nonumber 
\end{eqnarray}
where we have used the identity (\ref{tens_ident}) for the reference metric $\hat \gamma_{ij}$ in the last step.  We now insert this result into (\ref{div_T1}) to obtain
\begin{eqnarray} \label{div_T2}
&& \partial_t \left(  e^{6 \phi} \sqrt{\bar \gamma/\hat \gamma} \,S_i \right) + \Dflat_j (f_S)_i{}^j = 
\nonumber \\
&& ~~~~~~ \alpha e^{6\phi} \sqrt{\bar \gamma/\hat \gamma} \left( T_a{}^b \4\Gamma^a_{bi} - T_k{}^j \Gammaflat^k_{ji} \right),
\end{eqnarray}
where we have defined the momentum density as seen by a normal observer
\begin{equation} \label{S}
S_i \equiv \alpha T_i{}^t = \alpha g_{ic} T^{ct} 
= \alpha \rho_0 h u^t g_{ic} u^c = W^2 \rho_0 h v_i
\end{equation}
and its flux
\begin{eqnarray} \label{f_S}
(f_S)_i{}^j & \equiv & \alpha e^{6 \phi} \sqrt{\bar \gamma/\hat \gamma} \, T_i{}^j \\
& = & \alpha e^{6\phi} \sqrt{\bar \gamma/\hat \gamma} \,
(W^2 \rho_0 h v_i (v^j - \beta^j/\alpha) + p \delta_i{}^j) \nonumber
\end{eqnarray}
In the above manipulations we have used
\begin{equation}
g_{ic} u^c = W v_i.
\end{equation}

We now evaluate the source terms on the right-hand side of equation (\ref{div_T2})
\begin{equation}
T_a{}^b \4\Gamma^a_{bi} - T_k{}^j \Gammaflat^k_{ji} 
= T^{cb} \4\Gamma_{cbi} -  T^{cj} g_{kc}\Gammaflat^k_{ji} 
\end{equation}
by expanding the sums over the indices of $T^{ab}$ into terms that contain only the time component $T^{00}$, only mixed components $T^{0j}$, and only spatial components $T^{jk}$.   The time component picks up contributions from the spacetime connection symbol only,
\begin{equation} 
T^{00} \4\Gamma_{00i} 
= \frac{1}{2} T^{00} \partial_i g_{00} 
= \frac{1}{2}T^{00}  \partial_i (- \alpha^2 + \gamma_{jk} \beta^j \beta^k)
\end{equation}
Here the expression in parenthesis may be interpreted as a scalar on each spatial slice, so that we may replace the partial derivative $\partial_i$ with the covariant derivative $\Dflat_i$,
\begin{equation} \label{T00}
T^{00} g_{a0} \4\Gamma^a_{0i} = \frac{1}{2} T^{00} ( \beta^j \beta^k \Dflat_i \gamma_{jk} + 
2 \beta_k \Dflat_i \beta^k - 2 \alpha \Dflat_i \alpha).
\end{equation}
The mixed-components term may be written as
\begin{eqnarray} \label{T0j}
&& T^{0j}(\4\Gamma_{0ji} + \4\Gamma_{j0i} - \beta_k \Gammaflat^k_{ji} ) \nonumber \\
&& ~~~ = T^{0j} (\partial_i \beta_j - \beta_k \Gammaflat^k_{ji} ) = T^{0j} \Dflat_i \beta_j 
\nonumber \\
&& ~~~ = T^{0j} \Dflat_i (\gamma_{jk} \beta^k ) 
= T^{0j} (\gamma_{jk} \Dflat_i \beta^k + \beta^k \Dflat_i \gamma_{jk}) \nonumber \\
&& ~~~ = T^{0j} ( g_{jk} \Dflat_i \beta^k + \beta^k \Dflat_i \gamma_{jk}) \nonumber \\
&& ~~~ = T^0{}_k \Dflat_i \beta^k - T^{00} \beta_k \Dflat_i \beta^k + T^{0j} \beta^k \Dflat_i \gamma_{jk}.
\end{eqnarray}
We note that the middle term in the last line of (\ref{T0j}) will cancel the middle term in (\ref{T00}) when we add these expressions.  Finally, we evaluate the purely spatial components to find
\begin{eqnarray} \label{Tjk}
& & T^{jk}(\4\Gamma_{jki} - \gamma_{kl} \Gammaflat^l_{ji}) 
= T^{jk} (\Gamma_{jki} -\gamma_{kl} \Gammaflat^l_{ji}) \nonumber \\
& & ~~~ = T^{jk} \left( \frac{1}{2} \partial_i \gamma_{kj} - \gamma_{kl} \Gammaflat^l_{ji} \right) \nonumber \\
&& ~~~ =  T^{jk} e^{4\phi} \left( 2 \bar \gamma_{jk} \partial_i \phi  + \frac{1}{2} \partial_i \bar \gamma_{kj}
 - \bar \gamma_{kl} \Gammaflat^l_{ji} \right) \nonumber \\
 & & ~~~ =   T^{jk} e^{4 \phi}\left(2 \bar \gamma_{kj} \partial_i \phi + 
 \bar \gamma_{jl} ( \bar \Gamma^l_{ki} - \Gammaflat^l_{ki} ) \right) \nonumber \\
 & & ~~~ = T^{jk} e^{4 \phi} \left(2 \bar \gamma_{jk} \partial_i \phi + \frac{1}{2} \Dflat_i \bar \gamma_{jk} \right) \nonumber \\
 & & ~~~ =  \frac{1}{2} T^{jk} \Dflat_i \gamma_{jk},
\end{eqnarray}
where we have used equations (\ref{delta_Gamma}) and (\ref{delta_Gamma_compute}).  

Collecting terms we now define
\begin{eqnarray} \label{s_S}
 (s_S)_i & \equiv & \alpha e^{6\phi} \sqrt{\bar \gamma/\hat \gamma} \left( T_a{}^b \4\Gamma^a_{bi} - T_k{}^j \Gammaflat^k_{ji} \right) \\
 & = & \alpha e^{6 \phi} \sqrt{\bar \gamma/\hat \gamma} 
	\Big(  - T^{00} \alpha \partial_i \alpha + T^0{}_k \Dflat_i \beta^k  \nonumber \\
&& ~~~~~~	
	+ \frac{1}{2} \big( T^{00} \beta^j \beta^k   + 2 T^{0j} \beta^k + T^{jk} \big) \Dflat_i \gamma_{jk} 
	\Big) \nonumber
\end{eqnarray}
where, in a numerical calculation, $\Dflat_i \gamma_{jk}$ can be computed from
\begin{equation} \label{Dflat_gamma}
\Dflat_i \gamma_{jk} =  e^{4 \phi} \left(4\bar \gamma_{jk} \partial_i \phi + \Dflat_i \bar \gamma_{jk} \right).
\end{equation} 
Inserting the definition (\ref{s_S}) into (\ref{div_T2}) we obtain the Euler equation in the form
\begin{equation} \label{euler_cov}
\partial_t \left(  e^{6 \phi} \sqrt{\bar \gamma/\hat \gamma} \,S_i \right) + \Dflat_j (f_S)_i{}^j = (s_S)_i
\end{equation}
As for the continuity equation, this expression reduces to the corresponding Valencia form of the equation if a flat metric in Cartesian coordinates is chosen as the reference metric.  In a numerical application, we again express the covariant derivative in terms of partial derivatives and connection symbols, i.e.
\begin{eqnarray} \label{euler_partial}
& & \partial_t \left(  e^{6 \phi} \sqrt{\bar \gamma/\hat \gamma} \,S_i \right) + \partial_j (f_S)_i{}^j \\
&& ~~~~~ = (s_S)_i + (f_S)_k{}^j \Gammaflat^k_{ji} - (f_S)_i{}^k\Gammaflat^j_{kj}. \nonumber
\end{eqnarray} 

%==========================================================
\subsection{The energy equation}
\label{Sec:Energy}
%==========================================================

To derive an equation for the internal energy, we consider a projection along the normal $n_a$ of the conservation of energy-momentum (\ref{div_T}) and subtract the conservation of rest mass (\ref{continuity}),
\begin{equation}
n_a \nabla_b T^{ab} - \nabla_a (\rho_0 u^a) = 0,
\end{equation}
or 
\begin{equation}
\nabla_b (n_a T^{ab} + \rho_0 u^b) = T^{ab} \nabla_b n_a.
\end{equation}
On the left-hand side we again evaluate the divergence of a vector.  Proceeding exactly as in Section \ref{Sec:Continuity}, applying the identity (\ref{vec_ident}) once for the spacetime metric $g_{ab}$ and once for the reference metric $\hat \gamma_{ij}$, we arrive at the form
\begin{equation} \label{energy1}
\partial_t (e^{6 \phi} \sqrt{\bar \gamma/\hat \gamma}\, \tau)
+ \Dflat_j (f_\tau)^j =  - \alpha e^{6 \phi} \sqrt{\bar \gamma/\hat \gamma} \, T^{ab} \nabla_b n_a,
\end{equation}
where we have defined the internal energy as observed by a normal observer
\begin{equation} \label{tau}
\tau \equiv W^2 \rho_0 h - p - D
\end{equation}
and the corresponding flux
\begin{equation} \label{f_tau}
(f_\tau)^j \equiv \alpha e^{6\phi} \sqrt{\bar \gamma/\hat \gamma} \, \left( \tau (v^j - \beta^j/\alpha) + pv^j \right).
\end{equation}
To evaluate the right-hand side we use both (\ref{spatial_metric_def}) and (\ref{K_def})
\begin{eqnarray}
T^{ab} \nabla_a n_b & = &
T^{ab} g_a{}^c g_b{}^d \nabla_c n_d \nonumber \\
& = & T^{ab} ( \gamma_a{}^c - n_a n^c)(\gamma_b{}^d - n_b n^d )\nabla_c n_d \nonumber \\
& = & T^{ab} (- K_{ab} - \gamma_b{}^c n_a \partial_c \ln \alpha),
\end{eqnarray}
where the last term contains the acceleration of the normal observer
\begin{equation}
a_a \equiv n^b \nabla_b n_a = \gamma_a{}^b \partial_b \ln \alpha.
\end{equation}
We also expand 
\begin{eqnarray}
T^{ab} K_{ab} & = & T^{ab} g_{ac} g_{bd} K^{cd}  \\
& = & T^{00} \beta_i \beta_j K^{ij} + 2 T^{0i} \beta_j \gamma_{ik} K^{jk} + T^{jk} K_{jk}
\nonumber
\end{eqnarray}
and 
\begin{equation}
T^{ab} \gamma_b{}^c n_a \partial_c \ln \alpha = 
- T^{00} \beta^i \partial_i \alpha - T^{0i} \partial_i \alpha. 
\end{equation}
Collecting terms we define
\begin{eqnarray} \label{s_tau}
& & s_\tau \equiv \alpha e^{6\phi}\sqrt{\bar \gamma/\hat \gamma}
\Big(T^{00} (\beta^i \beta^j K_{ij} - \beta^i \partial_i \alpha) +  \nonumber \\
& & ~~~~~~~~~~~~ T^{0i}(2 \beta^j K_{ij} - \partial_i \alpha)
+ T^{ij} K_{ij} \Big)
\end{eqnarray}
and write equation (\ref{energy1}) as
\begin{equation} \label{energy_cov}
\partial_t (e^{6 \phi} \sqrt{\bar \gamma/\hat \gamma}\, \tau)
+ \Dflat_j (f_\tau)^j =  s_\tau.
\end{equation}
As for the continuity equation (\ref{D_cov}) this equation should be evaluated numerically by expanding the covariant derivative into a partial derivative and connection symbols,
\begin{equation} \label{energy_partial}
\partial_t (e^{6 \phi} \sqrt{\bar \gamma/\hat \gamma}\, \tau)
+ \partial_j (f_\tau)^j =  s_\tau - (f_\tau)^k \Gammaflat^j_{jk}.
\end{equation}

%==========================================================
\subsection{The generalized Valencia formulation}
\label{Sec:hydro_summary}
%==========================================================

The continuity, Euler and energy equations can be cast in a compact form by 
combining the conservative variables $D$, $S_i$ and $\tau$, given by equations (\ref{D}), (\ref{S}) and (\ref{tau}),  into a vector
\begin{equation}
\vec{q} = e^{6\phi} \sqrt{\bar \gamma/\hat \gamma} \, (D, S_i,\tau).
\end{equation}
We also define a corresponding flux vector
\begin{equation}
\vec{f}^{(j)} = \left( (f_D)^j, (f_S)_i{}^j, (f_\tau)^j \right)
\end{equation}
from equations (\ref{f_D}), (\ref{f_S}) and (\ref{f_tau}), as well as a source vector
\begin{equation}
\vec{s} = ( 0, (s_S)_i, s_\tau)
\end{equation}
from equations (\ref{s_S}) and (\ref{s_tau}).  The continuity equation (\ref{D_cov}), the Euler equation (\ref{euler_cov}) and the energy equation (\ref{energy_cov}) can then be combined into a single equation
\begin{equation}
\partial_t \vec{q} + \Dflat_j \vec{f}^{(j)} = \vec{s}.
\end{equation}
As expected, this flux-conservative form of the equations is in complete analogy to that of the original Valencia formulation.  The latter can be recovered by choosing the reference metric to be the flat metric in Cartesian coordinates, so that $\sqrt{\hat \gamma} = 1$ and $\Dflat_i = \partial_i$.  Reversing the process, our equations can be obtained from the original Valencia formulation by (a) dividing every determinant of the metric by that of the reference metric, and (b) replacing every spatial partial derivative, both in the flux terms and the source terms, with covariant derivatives with respect to the reference metric.  

%==========================================================
\subsection{Comparison with the original Valencia formulation}
\label{Sec:discussion}
%==========================================================

Before experimenting with our reference-metric formulation in numerical simulations in spherical polar coordinates in Sections \ref{Sec:SPC} and \ref{Sec:examples}, it is useful to compare some of its more general features with that of the original Valencia formalism.

We first note that the equations of relativistic hydrodynamics, when expressed in a reference-metric approach, mesh well with the equations for the gravitational fields, if they are also expressed with the help of a reference metric.  For example, the covariant derivatives of the conformal metric $\Dflat_i \bar \gamma_{jk}$ that appear in the flux term (\ref{s_S})  are also used to compute the $\Delta \Gamma^i_{jk}$ in equation (\ref{delta_Gamma_compute}).

Another attractive feature of our formalism is that, in the reference-metric approach, all conserved variables, fluxes and source terms are defined as spatial, tensorial quantities. In the original formulation, on the other hand, these quantities transform as tensor densities with non-zero weight.

We can also anticipate an important advantage of our formalism in
numerical applications.   For simplicity, consider a static and
spherically symmetric star, for which the momentum densities vanish,
$S_i = 0$, and for which $D$ and $\tau$ depend on the radius $r$ only.
We also assume $\beta^i = 0$, that the spatial metric is expressed as $\gamma_{ij} = e^{4 \phi}
 \eta_{ij}$, where $\eta_{ij}$ is the flat metric in spherical polar coordinates,
 \begin{equation}
 \eta_{ij} = \mbox{diag}(1, r^2, r^2\sin^2\theta),
 \end{equation}
and where $\phi$, as well as the lapse function $\alpha$, depend on $r$ only.   Clearly we would like the momentum densities to remain zero, $\partial_t S_i = 0$.  It is instructive to evaluate the $\theta$-component of this equation in both the original and the generalized Valencia formulation.

For the original Valencia formulation, we consider equation (\ref{euler_cov}) with $\hat \gamma = 1$ and $\Dflat_i = \partial_i$.  The flux term (\ref{f_S}) can then be written
\begin{equation}
(f_S)_i{}^j = \alpha \sqrt{\gamma} \, p \, \delta_i{}^j,
\end{equation}
where $\sqrt{\gamma} = e^{6\phi} \sqrt{\bar \gamma} = e^{6\phi} r^2\sin \theta$ in spherical symmetry .  Inserting this term into (\ref{euler_cov}) we obtain for the $\theta$-component
\begin{equation} \label{euler_theta_1}
\partial_j (f_S)_\theta{}^j = \alpha p \,\partial_\theta \sqrt{\gamma},
\end{equation}
which is non-zero.  Analytically, this term is canceled exactly by the term
\begin{eqnarray} \label{euler_theta_2}
(s_S)_\theta & = & \frac{\alpha \sqrt{\gamma}}{2}  T^{jk} \partial_\theta \gamma_{jk} = 
	\frac{\alpha \sqrt{\gamma}}{2}  p \, \gamma^{jk} \partial_\theta \gamma_{jk} =
 \nonumber \\
 & = & \alpha p \, \partial_\theta \sqrt{\gamma}
\end{eqnarray}
in the source term (\ref{s_S}).  Here we have used the identity 7.7 (d) of \cite{LigPPT75} in the last step.  Numerically, however, the two terms (\ref{euler_theta_1}) and (\ref{euler_theta_2}) are treated very differently.  In an HRSC scheme, the term (\ref{euler_theta_1}) is evaluated from a derivative of the fluxes at the cell interfaces, which are computed from a suitable reconstruction method.  The source term (\ref{euler_theta_2}), on the other hand, is computed at the cell centers.  Therefore, the two terms do not cancel exactly.  We have confirmed in our numerical simulations that the resulting numerical error leads to an increasingly large momentum density $S_\theta$ which breaks spherical symmetry and ultimately spoils the numerical simulation.

In our generalized formulation, on the other hand, both the flux and source terms vanish individually.  The flux term (\ref{f_S}) is now
\begin{equation}
(f_S)_i{}^j = \alpha \sqrt{\gamma/\hat \gamma} \, p \delta_i{}^j = \alpha e^{6 \phi} p \delta_i{}^j,
\end{equation}
which no longer depends on $\theta$.   We then have
\begin{eqnarray}
\Dflat_j (f_S)_{\theta}{}^j & = & \partial_j (f_S)_\theta{}^j + (f_S)_\theta{}^k \Gammaflat^j_{kj}
 - (f_S)_k{}^j \Gammaflat^k_{\theta j} \\
 & = & \partial_\theta (\alpha e^{6\phi} p) + 
 \alpha e^{6\phi} p \left( \Gammaflat^j_{\theta j} - \Gammaflat^j_{\theta j} \right) = 0. \nonumber
\end{eqnarray}
The source term $(s_S)_\theta$ also vanishes identically since we now replace $\partial_\theta$ with $\Dflat_\theta$ in (\ref{euler_theta_2}).  Using (\ref{Dflat_gamma}) we have
\begin{equation} 
\Dflat_\theta \gamma_{jk} = e^{4 \phi} \Dflat_\theta \eta_{jk} = 0.
\end{equation}
As a consequence, the generalized formalism no longer relies on a numerical cancellation between flux and source terms.  We have found that this makes a dramatic difference in numerical simulations, as we will describe in Section \ref{Sec:examples} below.

This problem has been recognized before, of course. In general
relativistic hydrodynamics this issue has been addressed by
\cite{Call10,NeiC00}. In particular,~\cite{Call10} presented a
generalization of the general relativistic hydrodynamics equations to
handle this pressure term in a similar fashion. In the simulations of \cite{Dim01,Dur06,Mue09}, which adopt spherical polar
coordinates, a factor of $r^2 \sin \theta$ is factored out from at
least some terms in the Euler equation. This approach is also
implemented in some versions of the CoCoNut code \cite{coconut}.  In
fact, the same issues arise in Newtonian hydrodynamics, and similar
solutions have been used in Newtonian simulations~\cite{AndPR68}.  Our approach is more general in that it allows for an (almost) arbitrary reference metric, and it goes beyond just factoring out one term, in that it treats all terms as tensorial objects in a reference-metric framework.  The resulting formalism has all the advantages that we describe above.

%==========================================================
\section{Numerical implementation in spherical polar coordinates}
\label{Sec:SPC}
%==========================================================

\subsection{BSSN equations in covariant form}
\label{Sec:bssn}

 In spherical polar coordinates, the evolution of the gravitational fields can be accomplished by adopting the BSSN formalism  \cite{NakOK87,ShiN95,BauS98} in a covariant, reference-metric  approach \cite{Bro09}, and  by using a PIRK time integration method \cite{BauMCM13,MonC12,CorCD12} that handles the coordinate singularities very effectively (these singularities appear both at the origin, where $r = 0$, and on the axis where $\sin \theta = 0$). 

We note that an additional challenge is that inverse factors of $r$
and $\sin \theta$ appear through the dynamical variables themselves,
and it is therefore important to treat these appearances of $r$ and
$\sin \theta$ analytically~\cite{BauMCM13}.  In the implementation used in this paper we 
represent all tensorial quantities in an
orthonormal frame so that the correct powers of $r$ and $\sin\theta$
are absorbed in the unit vectors, as suggested in footnote 2
of~\cite{BauMCM13}. In addition to the spatial conformal
metric $\bar \gamma_{ij}$ and the conformal factor exponent $\phi$, the
BSSN equations evolve the trace of the extrinsic curvature, $K$, the
conformal trace-less part of the extrinsic curvature,  $\bar A_{ij}$,
and the vector $\bar \Lambda^i$ that plays the role of the ``conformal
connection functions" $\bar \Gamma^i$ in the original BSSN
formulation. We refer to~\cite{BauMCM13} for the explicit form of the BSSN equations that is implemented in the numerical code.

Before the BSSN equations can be integrated, we have to specify
coordinate conditions for the lapse $\alpha$ and the shift
$\beta^i$. We will adopt a ``non-advective" version of what has become
the ``standard gauge" in numerical relativity codes using the BSSN formulation.  Specifically, in all dynamical spacetime simulations we use the ``1+log" condition for the lapse \cite{BonMSS95} in the form
\begin{equation} \label{1+log}
\partial_t \alpha = - 2 \alpha K,
\end{equation}
and the ``Gamma-driver" condition for the shift \cite{AlcBDKPST03} in the form
\begin{subequations} \label{gammadriver}
\begin{eqnarray}
\partial_t \beta^i & = & B^i \\
\partial_t B^i & = & \frac{3}{4} \partial_t \bar \Lambda^i,
\end{eqnarray}
\end{subequations}
where $B^i$ is an auxiliary vectorial quantity.

\subsection{Time integration}
\label{Sec:pirk}

The code uses a second-order PIRK method to integrate the evolution equations in time. The PIRK scheme is applied to the hydrodynamic and BSSN evolution equations as follows.  Firstly, the hydrodynamic conserved quantities, the conformal metric components $\bar \gamma_{ij}$, the conformal factor $\phi$, the lapse function $\alpha$ and the shift vector $\beta^i$ are evolved explicitly; secondly, the traceless part of the extrinsic curvature, $\bar A_{ij}$, and the trace of the extrinsic curvature $K$ are evolved partially implicitly, using updated values of $\alpha$, $\beta^i$, $\phi$ and $\bar \gamma_{ij}$; then, the $\bar \Lambda^{i}$ are evolved partially implicitly, using the updated values of $\alpha$, $\beta^i$, $\phi$, $\bar \gamma_{ij}$, $\bar A_{ij}$ and $K$. Finally, $B^i$ is evolved partially implicitly, using the updated values of the previous quantities.  Lie derivative terms and matter source terms are always included in the explicitly treated parts. We refer to Appendix B in~\cite{BauMCM13} for the expressions of the source terms included in the PIRK operators.

We have implemented  two versions of the reference-metric approach to
the general relativistic hydrodynamic equations. In the first version,
which we call the {\it full approach}, we apply the reference-metric
approach to all five equatitions, that is the continuity equation
(\ref{D_partial}), the Euler equation (\ref{euler_partial}) and the
energy equation (\ref{energy_partial}).  In an alternative {\it
  partial approach}, we apply the reference-metric approach only to
the Euler equation (\ref{euler_partial}) while the continuity equation
and the energy equation are left in the original Valencia form.  We note that the partial approach casts the equations in a form that
  is closer to the modifications proposed by ~\cite{Call10,NeiC00} than the full approach.

\subsection{Numerics}
\label{Sec:numerics}

We adopt a cell-centered grid.  Specifically, we divide the physical domain covered by our grid, $0 < r < r_{\rm max}$, $0 < \theta < \pi/2$ and $0 < \varphi < 2 \pi$ into $N_r \times N_\theta \times N_\varphi$ cells with uniform coordinate size
\begin{equation}
\Delta r = r_{\rm max}/ N_r,~~~~
\Delta \theta = \pi/ 2 N_\theta,~~~~
\Delta \varphi = 2 \pi/ N_\varphi.
\end{equation}
We refer to Fig.~1 in~\cite{BauMCM13} for a schematic representation of our cell-centered grid structure in spherical polar coordinates (note, however, that we adopt equatorial symmetry here, while no symmetry condition was adopted in \cite{BauMCM13}). Because of our fourth-order finite differencing scheme we need to pad the interior grid with three layers of ghost zones. Except at the outer boundary, each ghost zone corresponds to some other zone in the interior of the grid (with some other value of $\theta$ and $\varphi$), so that these ghosts zones can be filled by copying the corresponding values from interior grid points. We again refer to \cite{BauMCM13} for a more detailed discussion.

For the solution of the BSSN equations we adopt a centered, fourth-order finite differencing representation of the spatial derivatives.  For each grid point, the finite-differencing stencil therefore involves the two nearest neighbors in each direction.  An exception from our centered, fourth-order differencing are advective derivatives along the shift, for which we use a fourth-order (one-sided) upwind scheme. At the outer boundary we also require two ghost zones. We impose a Sommerfeld boundary condition, which is an approximate implementation of an outgoing wave boundary condition, to fill these ghost zones.  We also adopt equatorial plane reflection symmetry conditions to reduce the computational cost of the simulations but we note that our code can run without this assumption.  As in \cite{BauMCM13} we use Kreiss-Oliger~\cite{KreO73} dissipation to suppress the appearance of high frequency noise at late times.

We use a HRSC scheme to solve the general relativistic hydrodynamic equations. In particular, we have implemented a second-order slope limiter reconstruction scheme, the MC limiter \cite{Van77}, to obtain the left and right states of the primitive hydrodynammic variables at each cell interface, and the HLLE approximate Riemann solver~\citep{HarLL83,Ein88}.

An important ingredient in numerical simulations based on finite difference schemes to solve the hydrodynamic equations is the treatment of vacuum regions. The standard approach is to add an atmosphere of very low density filling these regions~\cite{Fon02}. We follow this approach and treat the atmosphere as a perfect fluid with a rest-mass density several orders of magnitude smaller than that of the bulk matter. The hydrodynamic equations are solved in the atmosphere region as in the region of the bulk matter. If the rest-mass density $\rho$ or specific internal energy $\epsilon$ fall below the value set for the atmosphere, these values are reset to have the atmosphere value of the respective primitive variables.

Unless stated otherwise we adopt a $\Gamma$-law equation of state
\begin{equation}
P=\left(\Gamma -1\right)\rho\epsilon,
\end{equation}
where $\Gamma=1+1/N$ and $N$ is the polytropic index.

%==========================================================
\section{Numerical examples}
\label{Sec:examples}
%==========================================================

We consider a number of test cases to demonstrate that it is possible
to obtain stable and robust general relativistic hydrodynamic
evolutions using spherical polar coordinates following the
reference-metric approach.  Although the initial data we consider are either spherically or axially symmetric we do not apply any symmetry condition except for the equatorial reflection symmetry.  In Section \ref{Sec:Cowling} we follow the common approach of keeping the spacetime fixed during the numerical evolution (known as the Cowling approximation \cite{Cow41}) in order to assess the hydrodynamical evolution independently from the spacetime evolution. In Section \ref{Sec:Dynamical} we relax this approximation and present several tests in dynamical spacetimes, including collapse to black holes.  We believe that our results represent the first successful, self-consistent general relativistic hydrodynamics simulations in spherical polar coordinates, without the need for a regularization or symmetry assumptions.
 
%Fig 1
\begin{figure}
\includegraphics[angle=0,width=8.5cm]{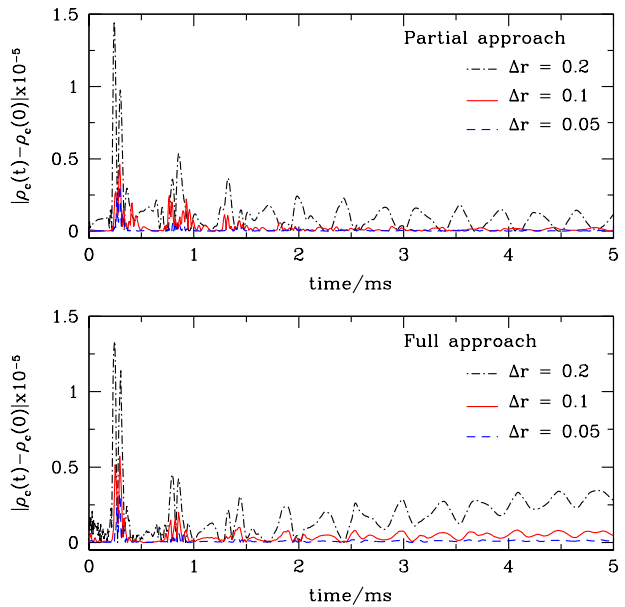}
\caption{Time evolution of the difference $|\rho_c(t)-\rho_c(0)|$ for
  the spherical relativistic star in the Cowling
  approximation, for both the full (lower panel) and partial
  approach (upper panel) using three different resolutions in the
  radial direction such that the grid spacing varies as $\Delta
  r=0.2, 0.1, 0.05$.  The full approach produces noisier results
  initially, and leads to a larger drift in the long term evolution of
  the central rest-mass density than the partial approach. Overall, the error
  decreases with increasing resolution in both approaches.}  
\label{fig1}
\end{figure}

%Fig 2
\begin{figure}
\includegraphics[angle=0,width=8.5cm]{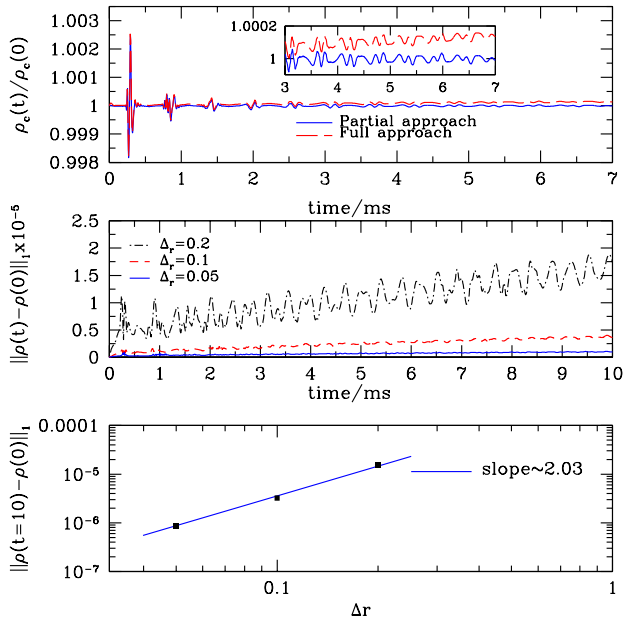}
\caption{We show in the upper panel the time evolution of the
  normalized central density for the spherical relativistic star using
  a grid spacing of $\Delta  r=0.05$ for both the full (dashed line) and partial
  approach (solid line). As we also saw for coarser grids,
  the drift in the time evolution of the central density for the full
  approach is larger than for the partial approach. The middle panel
  displays the time evolution of the L1-norm $||\rho(t)-\rho(0)||_{1}$
  computed inside the star for different resolutions for the partial approach. Finally, we show in
  the lower panel the convergence rate of the L1-norm
  $||\rho(t)-\rho(0)||_{1}$ at $t=5$ ms is approximately $2.03$
  for the partial approach.}  
\label{fig2}
\end{figure}

\subsection{Fixed spacetime evolutions}
\label{Sec:Cowling}

\subsubsection{Spherical stars}
\label{Sec:tov_cowling}

As a first test we consider a non-rotating relativistic star. The
initial data for the fluid, as well as the fixed spacetime geometry,
are given by the solution of the Tolman-Oppenheimer-Volkoff (TOV)
equations \cite{Tol39,OppV39}.  We focus on a polytropic TOV star with
$\Gamma=2$, and with a gravitational mass of about 85\% of the
maximum-allowed mass.  For this model, the central density is about
40\% of that of the maximum mass model.  In our code
units, for which $M_{\odot} = 1$, the gravitational mass of this star is $M = 1.4$
and the central density is $\rho_c = 1.28 \times{-3}$.  We adopt a numerical grid of
size $(100N,2,2)$ with $N=1,2,4$ and place the outer boundary at
$r_{\rm max} = 20$, which equal approximately two times the radius of
the star. We evolve the fluid using both the full and partial approach, as discussed in Section~\ref{Sec:pirk}.

In Fig.~\ref{fig1} we show the time evolution of the difference
$|\rho_c(t)-\rho_c(0)|$ for both approaches.  The truncation errors
resulting from the finite difference representation of the PDEs excite
small periodic radial oscillations which manifest themselves  as
periodic variations of the hydrodynamical quantities with respect to
their initial values. We obtain convergence of the numerical results
with increasing resolution with both approaches. However, we observe
that the initial phase is noisier in the full approach than in the
partial approach, and also that there is a larger drift in the long
term evolution of the rest-mass density in the full approach (see also
the upper panel in Fig.~\ref{fig2} which displays the time evolution
of the normalized central density using a grid spacing of $\Delta
r=0.05$ for both approaches).  We believe that these differences are
caused by the presence of source terms in the reference-metric version of
the continuity and energy equations; moreover, these source terms
contain singular terms that scale, e.g., with $1/r$.  These
source terms increase the truncation error in the evaluation of the
right-hand-side of the continuity equation for $r\simeq 0$. In fact,
in the``full approach'' approach, the continuity equation is written
as a ``balance law'' rather than as a  ``conservation law''
(e.g~\cite{Lev92}). While our PIRK scheme is able to handle these
singular terms in a stable fashion, they do lead to a larger numerical
error than that found in the evolution with the partial approach
(we note that the partial approach is closer to the modifications
proposed by ~\cite{Call10,NeiC00} than the full approach). We also
observe that the numerical error associated with the full approach is
larger for axisymmetric fluid configurations. We therefore adopt the
partial approach for the remainder of the paper.   We also stress that
using the original version of the Euler equation leads to much larger
errors, and a violation of spherical symmetry (see Section
\ref{Sec:discussion}) that makes the code crash after a short time.   It is therefore crucial to accommodate the spherical polar coordinates in the Euler equation in some way.  We have found that the reference-metric formulation provides a both elegant and effective approach to handling this issue.

  The middle panel of Fig.~\ref{fig2} displays the time evolution of
  the L1-norm $||\rho(t)-\rho(0)||_{1}$ computed inside the star for
  the partial approach. We define the L1-norm of a function $f(t)$ as
  \begin{equation}
||f(t)||_{1}=\frac{1}{N_T}\sum_{i=1}^{N_{T}} |f_{i}(t)|,
\end{equation}
where $N_T$ is the total number of grid points inside the star. We
plot the L1-norm for different resolutions showing that the error
decreases with increasing resolution. We also observe that the
truncation errors at higher resolutions lead to smaller oscillations,
and that the damping of the periodic  oscillations remains small
during the entire evolution, which  highlights the low numerical
viscosity of the implemented  scheme. Finally, we show in the lower
panel that the convergence  rate of the L1-norm
$||\rho(t)-\rho(0)||_{1}$ at $t=5$ ms is approximately $2.03$. While the order of convergence of HRSC schemes reduces to first order at the stellar center and surface, the convergence of $||\rho(t) - \rho(0)||_1$ appears to be dominated by the higher-order convergence in the bulk of the star in this case.

%Fig 3
\begin{figure}
\includegraphics[angle=0,width=8.5cm]{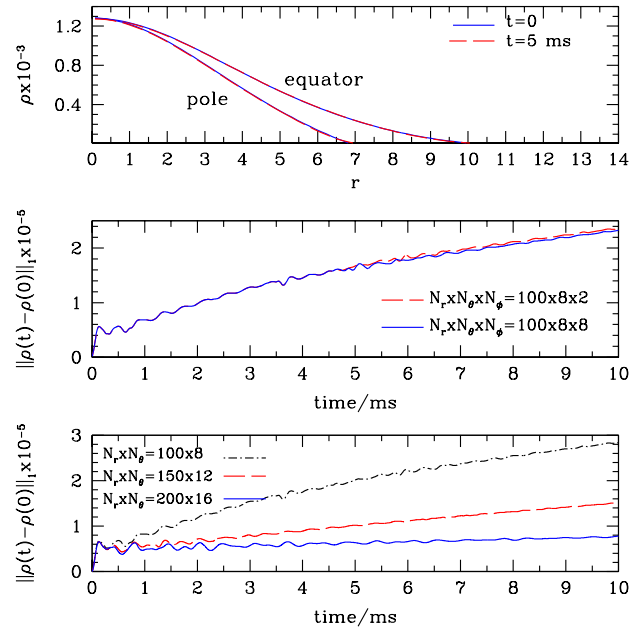}
\caption{Upper panel: Snapshots of the rest-mass density $\rho$ at the initial time $t=0$ 
  and at a later time $t=5$ ms for the evolution of a uniformly rotating
  star in the Cowling approximation.  We show profiles along one ray very close to the
  equator, and another close to the pole. Both profiles remain
  very similar to their initial data throughout the evolution.  Middle panel:
  the L1-norm $||\rho(t)-\rho(0)||_{1}$ for
  two simulations with $N_r=100$, $N_\theta=8$, and $N_{\varphi}=2,8$.  Lower panel:  the
  L1-norm $||\rho(t)-\rho(0)||_{1}$ for three simulations performed
  with grids consisting of $(100,8,2)$, $(150,12,2)$ and $(200,16,2)$
  points, respectively.}
\label{fig3}
\end{figure}

\subsubsection{Rotating stars}
\label{Sec:rotstar_cowling}

The numerical evolution of a rapidly rotating relativistic star is a more demanding test than the previous one, as it involves axisymmetric initial data in the strong gravity regime.  The initial data used for this test are the numerical solution of a stationary and axisymmetric equilibrium model of a rapidly and uniformly rotating relativistic star~\cite{BonGSM93}, which is computed using the Lorene code \cite{lorene}.

We consider a uniformly rotating star with the same $\Gamma=2$
polytropic equation of state as for the non-rotating model of
Sect.~\ref{Sec:tov_cowling}.  Our particular model has the same
central rest-mass density as the non-rotating model, but rotates at
$95\%$ of the mass-shedding limit (for a star of that central density); the corresponding spin period is approximately $0.7$ ms. The ratio of the polar to equatorial coordinate radii for this model is $0.67$.

For this test we adopt four grids of sizes
$(100,8,2)$, $(100,8,8)$, $(150,12,2)$ and $(200,16,2)$, and impose the
outer boundary at $30$, which equals approximately  three times the
equatorial radius.  In Fig.~\ref{fig3} (upper panel) we show the
initial and late-time profiles of the rest-mass density $\rho$, both
in a direction close to the equator and close to the axis.  Evidently,
these remain very close to their initial values throughout the
evolution, as they should,  and confirm the long term stability of the
simulation. The middle panel displays the L1-norm
$||\rho(t)-\rho(0)||_{1}$ for
two simulations with $N_r=100$, $N_\theta=8$, and $N_{\varphi}=2,8$. We note that even in the case
for only $N_{\varphi}=2$,  the two grid points in the
$\varphi$-direction belong to the
computational domain where the hydrodynamic equations are actually
evolved, and do not represent ghostzones. We see that the error is
almost the same independently of $N_{\varphi}$ for such axisymmetric
configuration and small differences only show up at late-times. Such
behavior highlights one of the advantages of using a coordinate system
well adapted to the geometry of the fluid configuration.  In the lower
panel of Fig.\ref{fig3}, we show the time evolution of the L1-norm
$||\rho(t)-\rho(0)||_{1}$ computed inside the star for three grids of
sizes $(100,8,2)$, $(150,12,2)$, and $(200,16,2)$,
respectively, demonstrating that the error decreases with increasing resolution. While, at late times, the errors decrease with increasing resolution, some of the perturbations at early times are triggered by numerical error originating at the stellar surface, where some of the fluid and spacetime variables are either discontinuous or have discontinuous derivatives.   As expected, these errors to not converge at the same rate as those for smooth functions.  We note that the order of convergence of the HRSC scheme
reduces to first order both at the center of the star and at its surface. In
addition, we use an static atmosphere which is not corotating with the
star (therefore inducing a larger error than in the test of a
spherical non-rotating star). The treatment of the
interface between the fluid configuration and the vacuum region is
one of the most challenging aspects for hydrodynamic codes using HRSC
schemes;  we refer to~\cite{RadRG13} for a recent discussion.

\subsection{Dynamical spacetime evolutions}
\label{Sec:Dynamical}

%Fig 4
\begin{figure}
\includegraphics[angle=0,width=8.5cm]{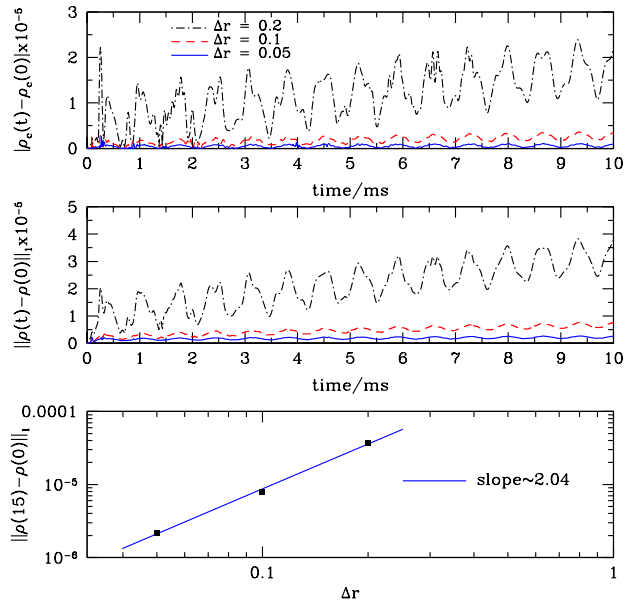}
\caption{We show in the upper panel the time evolution of the difference $|\rho_c(t)-\rho_c(0)|$ for
  the TOV model in a dynamical spacetime, using three different resolutions in 
  radial direction such that the grid spacing varies as $\Delta
  r=0.2, 0.1, 0.05$. The middle panel graphs the time evolution of the L1-norm
  $||\rho(t)-\rho(0)||_{1}$ computed inside the star, and the lower
  panel shows that the convergence rate of the L1-norm
  $||\rho(t)-\rho(0)||_{1}$ at $t=15$ ms is approximately $2.04$.}
\label{fig4}
\end{figure}

%Fig 5
\begin{figure}
\includegraphics[angle=0,width=8.cm]{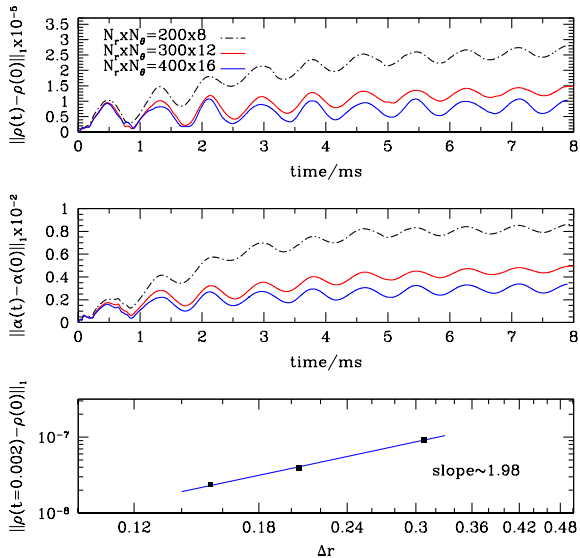}
\caption{ Dynamical spacetime evolution of a rotating relativistic
  star. The figure shows the time evolution of the L1-norm, computed
  inside the star,  of the rest-mass density $||\rho(t)-\rho(0)||_{1}$ (upper panel) and the
  lapse function $||\alpha(t)-\alpha(0)||_{1}$ (middle panel). For all
  simulations we impose the outer boundary at $r=60$ in our code
  units.  Error originating from the outer boundaries reaches the
  center at around $t =0.3$  ms, and triggers the oscillations visible
  in the graph.   As expected, the amplitude of the initial
  oscillation does not decrease with increasing resolution; however,
  for lower resolutions the amplitude continues to increase, while for
  higher resolutions it does not. We therefore measure the convergence
  rate before outer boundary and surface effects become the main
  source of error. We show (lower panel) that the convergence rate of the L1-norm
  $||\rho(t)-\rho(0)||_{1}$ at $t=0.002$ ms (for comparison, this corresponds to about
  one-tenth of the time needed by light to travel from the surface to
  the center, and approximately 1000 timesteps) is approximately $1.98$.}
\label{fig5}
\end{figure}

\subsubsection{Spherical stars}

As a first test of self-consistent evolutions of Einstein's equations
coupled to the equations of relativistic hydrodynamics we return to
the TOV solution. In particular, we use the same TOV star as in
Section \ref{Sec:tov_cowling}, but now we evolve the spacetime
dynamically rather than keeping it fixed.  We adopt the 1+log slicing
condition (\ref{1+log}) as well as the Gamma-driver shift condition
(\ref{gammadriver}). As in the Cowling tests, we choose a numerical
grid of size $(100N,2,2)$ with $N=1,2,4$ and place the outer boundary
at $r_{\rm max} = 20$, which equals approximately two times the radius of
the star, where we impose Sommerfeld boudary conditions for the
gravitational fields. 

  In Fig.~\ref{fig4} we show the time evolution of the difference
  $|\rho_c(t)-\rho_c(0)|$ using three different resolutions in the
  radial direction such that the grid spacing varies as $\Delta r=0.2,
  0.1, 0.05$.   As expected, the difference $|\rho_c(t) - \rho_c(0)|$ at
  decreases with increasing resolution. 
  The small value of the error demonstrates the ability of
  the code to maintain the equilibrium configuration. This is better
  shown in the middle panel where we plot the time evolution of the
  L1-norm $||\rho(t)-\rho(0)||_{1}$ computed inside the star, and in
  the lower panel that displays the L1-norm at a late time ($t=15$ ms)
  versus the radial grid spacing. The slope of approximately $2.04$ indicates
  that the convergence is second-order inside the star.

Finite-difference errors in the initial data trigger small amplitude radial pulsations of the star which are a sum of eigen modes of pulsation~\cite{Fon02}.  These finite-difference errors arise not only from  the hydrodynamic part of the code but also from the spacetime part that solves the full set of Einstein equations. It is expected that the star oscillates at the proper mode frequencies and therefore, it is possible to exploit this feature to check the consistency of the non-linear evolution by comparing numerical results for the stellar mode frequencies with the predictions from linear perturbation theory~\cite{FonSK00}. In fact, this has become an standard test for numerical relativity codes. The power spectral density of the maximum density  time evolution (for the grid with $(400,2,2)$ points) displays a peak for the fundamental mode at $\nu_{F}=1.427$ KHz and at $\nu_{H1}=3.945$ KHz for the first overtone. We find excellent agreement between our frequency peaks and the theoretical values~\cite{FonSK00,Fon02}; the relative errors for the two frequencies are less than 1\%.

\subsubsection{Rotating stars}

As a test that does not involve spherically symmetric initial data we
again consider relativistic rotating stars, but now evolve the spacetime
together with the fluid. We adopt the same model
as that in Section \ref{Sec:rotstar_cowling} and  three grids of sizes
$(200,8,2)$, $(300,12,2)$ and $(400,16,2)$, and
impose the outer boundary at $60$.  We therefore cover the rotating star by the same number of grid points as in Section
\ref{Sec:rotstar_cowling} while placing the outer boundary at
approximately six times the equatorial radius of the star. We notice that not only the interpolation of the
initial data from the Lorene computational domains onto our grid and
truncation errors due to the spacetime evolution, but in particular,
the outer boundary Sommerfeld condition for the gravitational fields induce oscillations
of larger amplitude than what we observed in the Cowling
approximation (where the Sommerfeld outer boundary condition does not
play any role as the gravitational fields do not evolve in time).  The oscillations are also visible in the spacetime
quantities. In Fig.~\ref{fig5} we plot the L1-norm
$||\rho(t)-\rho(0)||_{1}$ in the upper panel, and the L1-norm
$||\alpha(t)-\alpha(0)||_{1}$ in the middle panel, where both L1-norms
are computed inside the star.  Error originating from the outer boundaries reaches the
  center at around $t=0.3$  ms ($t=60$ in our code units), and triggers the oscillations visible
  in the graph.  As expected, the amplitude of the initial
  oscillation does not decrease with increasing resolution; however,
  for lower resolutions the amplitude continues to increase, while for
  higher resolutions it does not.  At very early times, the time evolution of
  the L1-norms shows that the error decreases with increasing
  resolution. In particular, we measure the convergence rate of the L1-norm
  $||\rho(t)-\rho(0)||_{1}$ at $t=0.002$ ms, well before the outer boundary
  conditions as well as the stellar surface (compare the discussion in Section \ref{Sec:rotstar_cowling}) affect the numerical evolution of the star.  In the lower panel of Fig.~\ref{fig5} we show that the convergence rate, at these early times, is
  approximately $1.98$. Most importantly, however, our results demonstrate that our
code can stably evolve rapidly rotating star for many dynamical
timescales.  

\subsubsection{Oppenheimer-Snyder collapse}

Oppenheimer-Snyder (OS) collapse is an analytical solution describing the collapse of a homogeneous dust sphere into a black hole~\cite{OppS39}.  This solution has served as a testbed for numerous numerical codes over the years.  Even though there is no complete analytical solution describing OS collapse in moving-puncture coordinates, several features of this solution can be obtained analytically (see \cite{Sta12}) and can be used to test our code. 
%Fig 6
\begin{figure}
\includegraphics[angle=0,width=8.cm]{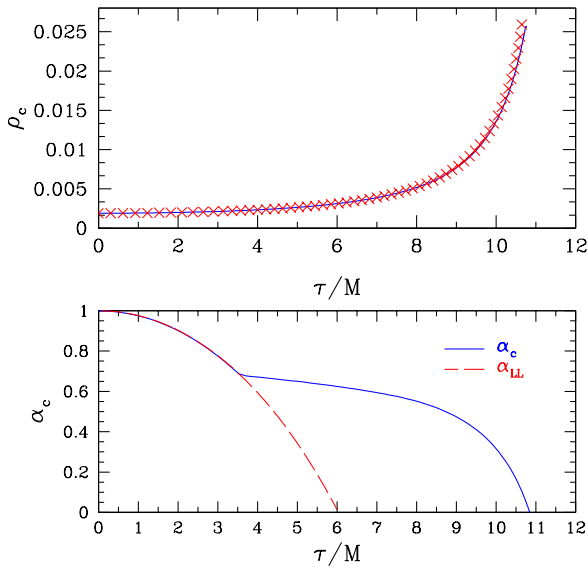}
\caption{Oppenheimer-Snyder collapse of a dust cloud to a black
  hole. In the upper panel we show the  time evolution of the central
  rest-mass density up to the approximate  time of black hole
  formation. The solid (blue) line is  the analytical solution for the
  central rest-mass density as a  function of the proper time, and the
  (red) crosses are the  numerical solution for the same quantity at a
  coordinate location  $r=0.1M$ (which avoids numerical artifacts that
  are the result of the  larger truncation errors caused the
  coordinate singularity at  $r=0$). The bottom panel shows the values
  of the lapse at the center, $\alpha_c$ (blue-solid), together with its lower limit, $\alpha_{\rm LL}$  (red-dashed), given by eq.~(\ref{lapse_LL}). }
\label{fig6}
\end{figure}

The initial data for OS collapse are obtained by writing the metric in isotropic coordinates. The exterior Schwarzschild metric then takes the form
\begin{equation}
dl^{2} = \left(1+\frac{M}{2r}\right)^4(dr^{2}+r^{2}d\Omega^2),
\end{equation}
where $dl^2$ denotes the spatial line element. The interior metric is
obtained by transforming the Friedmann metric to isotropic coordinates
and matching the conformal factors in the interior and exterior at the
surface of the star. The initial spatial line element then appears as
\begin{equation}
dl^2 = \psi^4(dr^2 + r^2 d\Omega^2)
\end{equation} 
with 
\begin{equation}
    \psi = \left\{\begin{array}{ll} 
   \displaystyle  \left( \frac{\left(1 + \sqrt{1 - 2M/R_0} \right)r_0 R_0^2}{2 r_0^3 + Mr^2}\right)^{1/2} \ ,  & r \le r_0 \ , \\
   \displaystyle  1 + \frac{M}{2r}  \ , & r > r_0 \ , 
    \end{array} \right.
\end{equation}
where 
\begin{equation}
r_0 = R_0 \left(1 - M/R_0 + \sqrt{1 - 2M/R_0}\right)/2
\end{equation}
(see \cite{Sta12}). The initial data also include $K_{ij} = 0$, $\beta^i=0$ and $\alpha=1$.  For our simulations here we choose the initial areal radius $R_{0}=5M$. The initial rest-mass density $\rho(0)$ is related to $R_0$ and the mass $M$ by
\begin{equation}
M=\frac{4\pi}{3}\rho(0)R_{0}^3.
\end{equation}

We evolve these initial data with moving-puncture coordinate conditions.   One gauge-invariant quantity that can be compared with the analytical solution is the central rest-mass density as a function of proper time.  In the upper panel of Fig.~\ref{fig6} we show this analytical solution as a solid (blue) line, and our numerical solution as (red) crosses.   Since the coordinate singularity at the $r=0$ leads to a relatively large truncation error at the center, we instead show numerical results for the rest-mass density at a the coordinate location$r=0.1M$.  We nevertheless find very good agreement between the numerical and analytical solution.
  
 As shown by \cite{Sta12}, in early stages of moving-puncture evolutions of Oppenheimer-Snyder collapse the lapse remains spatially constant in a region around the center.  This region is limited by a ``gauge wave" that originates at the surface and propagates toward the center.  Once this gauge wave reaches the center at a (proper) ``gauge time" $\tau_{\rm gauge}$, the region of spatially constant lapse disappears.   For $R_0 = 5M$, the gauge time is $\tau_{\rm gauge} \approx 3.54 M$.   Moreover, \cite{Sta12} show that (under conditions that generally hold) the central lapse $\alpha_c$ is greater or equal than a ``lower-limit" lapse $\alpha_{\rm LL}$ given by
\begin{equation} \label{lapse_LL}
\alpha_c \geq \alpha_{\rm LL}= 1+6\ln(a/a_{m}),
\end{equation}
where the scale factor $a$ is can be expressed parametrically as a function of proper time by
\begin{eqnarray}
a & = & \frac{1}{2} a_m ( 1 + \cos \eta ), \\
\tau & = & \frac{1}{2} a_m (\eta + \sin \eta),
\end{eqnarray}
and where the initial scale factor is given by
\begin{equation}
a_m = \left(\frac{R_0^3}{2M}\right)^\frac{1}{2}.
\end{equation}

In (\ref{lapse_LL}), equality holds as long as the lapse remains spatially constant at the center.  The arrival of the gauge wave at $\tau = \tau_{\rm gauge}$ marks a sudden departure of $\alpha_c$ from $\alpha_{\rm LL}$  (see also Fig.~2 in \cite{Sta12}).  Reproducing this behavior therefore serves as a stringent code test.

In the lower panel of Fig.~\ref{fig6} we show our numerical results for $\alpha_c$ together with $\alpha_{\rm LL}$ as given by (\ref{lapse_LL}).  As expected, we find excellent agreement between the two quantities at early times, and a sudden departure at $\tau \approx 3.5M$, very close to the theoretical value.  At late times, our simulation settles down to a Schwarzschild black hole in trumpet geometry \cite{HanHPBO06,HanHOBGS06,BauN07,Bro08,HanHO09}, which completes the collapse of the dust cloud to a black hole in moving-puncture coordinates.

\subsubsection{Collapse of a marginally stable spherical star to black hole}

%Fig 7
\begin{figure}
\includegraphics[angle=0,width=8.cm]{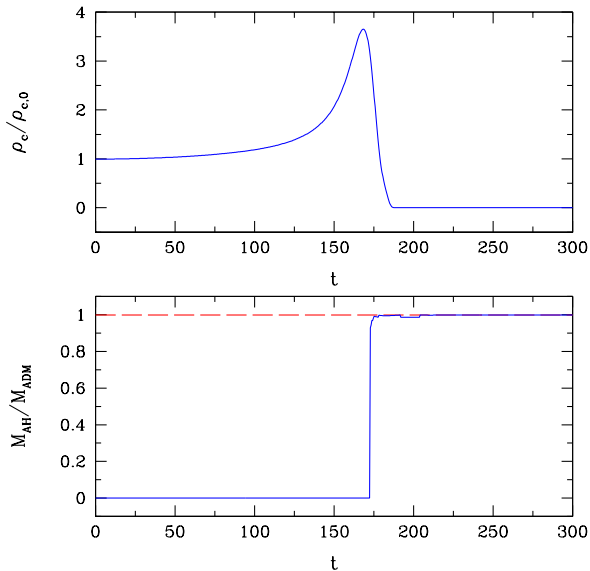}
\caption{Collapse of a marginally stable spherical star to a black hole.  In the upper panel we show
the time evolution of the normalized central density (measured at a coordinate radius $r=0.075$),  
and in the lower panel the apparent-horizon mass (solid line) in units of the ADM mass of the system (dashed line).}
\label{fig7}
\end{figure}

\begin{figure}
\includegraphics[angle=0,width=8.cm]{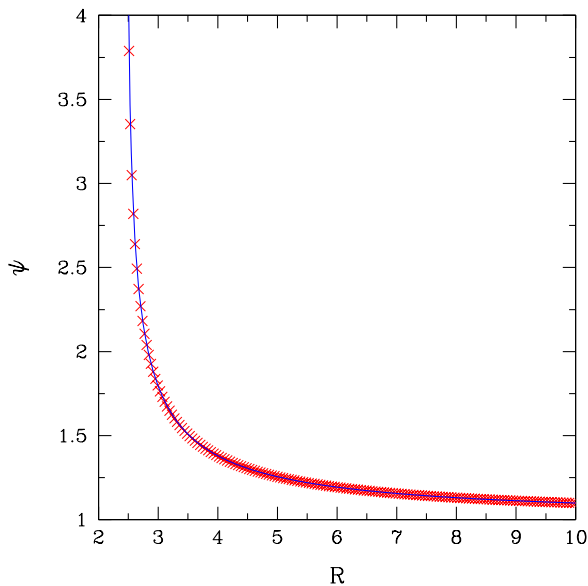}
\caption{Radial profile of the conformal factor $\psi$ at time $t = 300$ for the collapse of a marginally stable star to a black hole.  The (red) crosses mark our numerical results, while the (blue) line is the analytical solution for a maximally-sliced trumpet solution (see \cite{BauN07}).}
\label{fig8}
\end{figure}

We next test the capability of the code to follow black hole formation from the gravitational collapse of a marginally stable spherical relativistic  star. For this test, we consider a $\kappa=100$, $\Gamma=2$ polytropic star with central rest-mass density $\rho_{c}=3.15\times 10^{-3}$, so that its gravitational mass is $M=1.64$ and its baryon rest-mass $M_{*}=1.79$. In order to induce the collapse of the star, we initially decrease the pressure by 0.5$\%$.  We adopt moving-puncture gauge conditions, use a numerical grid of $(2000,2,2)$ points, and impose the outer boundary at~$r_{\rm max} = 100$.  We stopped the simulation at $t =300$ without encountering any instabilities.

In Fig.~\ref{fig7} we show the time evolution of the normalized central density (top panel) and the mass of the apparent horizon in units of the ADM mass of the system (bottom panel).  At early times the central density increases, reflecting the contraction of the collapsing star.  As an unambiguous signature of black-hole formation we first detect an apparent horizon at $t\sim 172$.  The mass of this horizon quickly settles down to the ADM mass of the spacetime; at $t=300$ the relative difference between the ADM mass and the horizon mass is approximately 0.2\%.  As discussed in detail by \cite{ThiBHBR11}, the gamma-driver shift condition (\ref{gammadriver}) leads to large grid stretching once a black hole forms; this effect leads to the decrease in the central density around the time of apparent-horizon formation that can be seen in the top panel of Fig.~\ref{fig7}.

In Fig.~\ref{fig8} we show a radial profile of the conformal factor $\psi$ at $t=300$.  Since we are using the ``non-advective" version of the 1+log slicing condition (\ref{1+log}), the evolution settles down to a Schwarzschild black hole in a maximally sliced trumpet geometry \cite{HanHOBGS06}.  This maximally sliced trumpet solution can be expressed analytically \cite{BauN07}, and is included as the solid (blue) line in Fig.~\ref{fig8}.  We find very good agreement.

The simulations of Oppenheimer-Snyder collapse in the previous Section and the collapse of a marginally stable star in this Section demonstrate that our implementation of relativistic hydrodynamics and of the gravitational fields can can accurately handle the transition between a regular spacetime (that of the star) and an irregular spacetime containing a puncture singularity at $r=0$.

%==========================================================
\section{Summary and discussion}
\label{Sec:summary}
%==========================================================

We derive and implement a reference-metric version of the equations of relativistic hydrodynamics.  Our equations are a generalization of the Valencia formulation \cite{BanFIMM97} and reduce to that when a flat metric in Cartesian coordinates is chosen as the reference metric.  They are expressed in flux-conservative form and allow for the implementation of HRSC methods.

The advantage of the reference-metric approach is that it provides a natural framework for curvilinear coordinate systems.  The resulting equations of hydrodynamics mesh well with those for the gravitational fields, when the latter are expressed in a reference-metric approach (see, e.g., \cite{Bro09,BauMCM13}).  Moreover, all conservative variables, fluxes and source terms are now defined as tensorial quantities.  We note that the induction equation for magnetic fields can be treated analogously, so that the equations of general relativistic magnetohydrodynamics can similarly be expressed in terms of a reference metric. 

Perhaps the most important property of our formalism is that it avoids certain numerical error terms that are present when the original Valencia formulation is implement in spherical polar coordinates, and which cause a deviation from spherical symmetry even for spherically symmetric initial data.  These problems are well known from both relativistic and Newtonian hydrodynamics simulations, and can alternatively be handled by factoring out geometric terms from the flux quantities.  Our approach is more general and goes further, in that it casts all terms in a consistent geometric framework.  

We implement two versions of this formalism in spherical polar coordinates.  In our ``full approach" we apply the reference-metric approach to all general relativistic hydrodynamic equations, while in a ``partial approach" we apply the reference-metric approach to the Euler equation only and leave the continuity and energy equations as given by the original Valencia formulation. We found that, although both approaches give reliable results, the second approach is more accurate and robust.  We have therefore adopted this partial approach in a number of tests, both in the Cowling approximation (in which the spacetime is kept fixed) and for dynamical spacetimes.  Specifically, we perform simulations of non-rotating and rotating relativistic stars, of Oppenheimer-Snyder collapse, and the collapse of a marginally stable spherical star.  Our code is capable of performing these numerical experiments, including collapse to black holes, with high accuracy.

To the best of our knowledge, we present the first stable and self-consistent general relativistic hydrodynamic simulations in dynamical spacetimes in spherical polar coordinates without the need of regularization or symmetry assumptions.   Many numerical codes of the traditional astrophysics community adopt spherical polar coordinates because they offer several advantages over Cartesian coordinates for simulations of single stars -- one important example are supernovae calculations.  Since, to date, methods for treating relativistic gravitational fields self-consistently had not been available in spherical polar coordinates, these codes rely on some approximate treatment of the gravitational fields.   Our results demonstrate that these approximations can be relaxed, and show how general relativistic hydrodynamics can be evolved self-consistently will fully dynamical gravitational fields in spherical polar coordinates.  We therefore believe that our methods offer a promising approach to implementing a self-consistent treatment of the gravitational fields in such existing codes, and we hope that they will prove to be useful in future relativistic astrophysics simulations.

\acknowledgments

PM thanks Scott Noble for valuable discussions. TWB gratefully acknowledges support from the Alexander-von-Humboldt Foundation and thanks the Max-Planck-Institut f\"ur Astrophysik for its hospitality.  This work was supported in part by the Deutsche Forschungsgemeinschaft (DFG) through its Transregional Center SFB/TR7 ``Gravitational Wave Astronomy'', and by NSF grant PHY-1063240 to Bowdoin College.

\bibliography{references}

\end{document}